\documentclass{article}

\usepackage{color}

\usepackage[english]{babel}

\usepackage[letterpaper,top=2cm,bottom=2cm,left=3cm,right=3cm,marginparwidth=1.75cm]{geometry}

\usepackage{amsmath}
\usepackage{graphicx}
\usepackage[colorlinks=true, allcolors=blue]{hyperref}
\usepackage{amssymb}

\author{
Sergey G. Rubin$^{a,b}$\thanks{e-mail: sergeirubin@list.ru}
        }

\date{\small\it
$^c$  National Research Nuclear University MEPhI (Moscow Engineering Physics Institute),\\ 
            Kashirskoe shosse 31, Moscow 115409, Russia \\
$^d$  N.I. Lobachevsky Institute of Mathematics and Mechanics,
	Kazan  Federal  University, \\
	Kremlyovskaya ulitsa 18,  Kazan 420008,  Russia}

\begin{document}

\title{Multidimensional arrow of time %Emergence of the Time Arrow \\ from Higher-Dimensional Expansion
}

\maketitle

\begin{abstract}
This paper investigates the influence of extra dimensions on the nature of the arrow of time. We demonstrate that the observed arrow of time can be explained by the monotonic growth of the multidimensional manifold's volume. Unlike traditional cosmological approaches based on the entropy of matter or radiation, our model identifies the primary temporal direction with the Bekenstein-Hawking-Wald entropy of the geometric background. 

By establishing a formal relation between the statistical weight of the multidimensional manifold and the multiplicity of causally disconnected regions, we reveal that time's directionality is driven by dominant entropy production in the higher-dimensional bulk. A key consequence of this approach is that the arrow of time remains a persistent feature for a 4D observer even in the vacuum limit. This global geometric evolution suppresses local statistical fluctuations and ensures a robust and stable entropy flow throughout the manifold.
\end{abstract}

\section{Introduction}

Observations reveal unambiguous irreversibility in the Universe, such as the formation of a star from a collapsing gas cloud—a process inherently accompanied by photon radiation and a definitive increase in entropy. All such phenomena are manifestations of the Second Law of Thermodynamics. While the variety of ''Arrows of Time'' and their explanations continue to provoke intense debate, the underlying origin of the subject remains elusive. Comprehensive reviews of the arrow of time across different scales and physical systems can be found in \cite{Ellis:2019mlk}, while the deep problem of universal arrow of time formation is also discussed in \cite{Rovelli:2019igm, AyalaOna:2023wwd}.

The behavior of any physical system is governed by three fundamental principles: Classical Motion, Quantum Laws, and the Growth of Entropy. The first two share a common origin in the Feynman path integral and serve as primary tools for precise system descriptions. The principle of entropy growth, however, is more universal; it applies to all systems but often yields less certain local predictions. A canonical example is a cloud of non-interacting particles expanding in a vacuum. In this case, neither classical nor quantum equations of motion dictate the macroscopic evolution of the cloud's boundary. Nevertheless, the Second Law of Thermodynamics provides a definitive prediction: the gas will expand into the surrounding vacuum rather than contract.

Entropy growth appears to be a necessary element for explaining the arrow of time. Simultaneously, causality lies at the foundation of physical processes and is closely related to the emergence of time's direction. For instance, the authors of \cite{Donoghue:2020mdd} discuss quantum causality in its relation to the arrow of time. In a seminal paper, Zeeman \cite{Zeeman1964CausalityIT} demonstrated that causality alone is sufficient to dictate the Lorentz group as the fundamental symmetry of spacetime.

Penrose suggested that a monotonically increasing time coordinate is not a fundamental background but rather a consequence of the Universe's progression toward states of higher gravitational entropy. This is encapsulated in the Weyl Curvature Hypothesis, implying that gravitational degrees of freedom are encoded in the geometry of space. Relatedly, Causal Set Theory \cite{Bombelli:1987aa} replaces the continuous spacetime of General Relativity with discrete structures, where coordinates are substituted by causal set elements and labels encoding causal relations. Furthermore, Energetic Causal Sets \cite{Cortes:2017smj} assign intrinsic energy and momentum to each event, propagating these quantities through causal processes.

At the same time, another widespread and promising branch of research is widely devoted to the idea of extra dimensions. The presence of extra dimensions could affect entropy growth even in the absence of matter fields and influence the arrow of time.

We consider a warped product metric defined on a higher-dimensional manifold, where our four-dimensional spacetime is embedded in a $D=4+n$-dimensional bulk. The expansion of the extra dimensions leads to a non-zero Wald entropy, which could be related to the gravitational entropy \cite{Hu:2021pfh, Chakraborty:2024rpi, Clifton:2013dha} associated with the Weyl tensor.

In this framework, we show that the entropy growth of the compact $n$-dimensional extra space, governed by a fundamental postulate of causality, inherently generates the observed arrow of time.

The paper is organized as follows: Section 2 introduces the fundamental elements of our model, including the statistical weight of geometry and the Postulate of Causality. In Section 3, we apply this framework to $f(R)$ multidimensional model and calculate the corresponding Wald entropy. Section 4 discusses the connection between the arrow of time and the expansion of the extra-dimensional bulk, followed by a discussion of the results and concluding remarks in Section 5.

\section{Necessary elements to produce the arrow of time}

The extra-dimensional arrow of time discussed herein is analogous to the well-known cosmological arrow of time. Both are predicated on {spatial} expansion and the concomitant growth of entropy. However, explaining the inevitability of entropic increase relies on the prior assumption of a time arrow, creating a circularity. This circular logic can be resolved by introducing a Postulate of Causality — an independent principle that does not follow from established physical laws. This is the {likely} reason {for the rising} activity in this direction, {beginning with} the works of Penrose \cite{Penrose1979}, Sorkin \cite{Bombelli:1987aa}, and Smolin \cite{Cortes:2014ola}. 

\subsection{Causality}%\label{Causality}

The  {concept} of a time-asymmetric layer of physical laws underlying general relativity and quantum theory was proposed by Roger Penrose in 1979 \cite{Penrose1979}.
This  {hypothesis} was further supported by Marina Cortes and Lee Smolin \cite{Cortes:2017smj}, who introduced a large number of discrete states $S = \{I, J, K, L, \ldots\}$ forming energetic causal sets governed by an evolution rule. In this model, each state $I$ has a unique successor $J$. A similar  {framework} is developed in the study of causal sets \cite{Bombelli:1987aa, Surya:2019ndm}, where discrete states $A, B, C, \ldots$ form a causal set $\mathcal{A}$, and the order $A \prec B \prec C \prec \ldots$ denotes a "precedes" causal relation.
This  {notion} is widely discussed in the literature (see, e.g., \cite{Rovelli:2019igm}). The physical essence of a ``state'' can be characterized by energy and momentum \cite{Cortes:2017smj}, the discrete structure of 3D space \cite{Dowker:2021zel}, or entanglement entropy \cite{Sorkin:2016pbz}. The fundamental role of entropy in this context was specifically investigated in \cite{albert2003time}.

The most known arrows of time — the thermodynamic, cosmological, and psychological — are deeply interconnected with the entropy. %It can be concluded that causality and entropy are intrinsically linked, together forming the foundation for the phenomenon of the arrow of time. 
While the intimate connection between the progression of time and the increase in entropy has long been evident, deriving the former from the latter remains a non-trivial task \cite{Kiefer:2009tq}. We  outline a mechanism where the arrow of time emerges from the entropic ordering of states in a fundamental causal set. This provides a potential pathway to explain how a temporal direction arises from a more primitive, ordered structure, addressing a key open problem in modern physics \cite{Ellis:2019mlk,Mukohyama:2013ew}.

Our starting point is the following

\bigskip

\textbf{Postulate of Causality}

\noindent \textit{There exists a directed causal chain of transitions, $A \prec B \prec C \prec \cdots$, between macro states $A,B,C,...$.  
Each transition  $\prec$ leads to a successor state with larger entropy, 
}

\bigskip

According to Penrose, such a postulate should be foundational to physics and, as such, should not rely on pre-existing physical concepts. In this sense, the Postulate of Causality involves neither an a priori definition of "time" nor the specific details of a "state."
Note that entropy does not "increase with time,"\, but rather grows along the causal chain of transitions by definition. This article investigates the relationship between this postulate and the empirical manifestation of the arrow of time.

The main aim of this study is to derive the presence of the arrow of time through an appropriate choice of the macrostates involved in the postulate. We base our work on the Bekenstein-Hawking-Wald entropy definition, which relates entropy to the number of microstates within the horizon.

The logic for the emergence of the arrow of time with the help of the Postulate of Causality is as follows:  
Choose a physical set of states $\{A\}$; define its entropy $S_A$; prove that the entropy is a monotonically increasing function along the temporal coordinate $t$; finally, apply the Postulate of Causality to show the arrow of time presence.
%According to the Postulate, a transition occurs when the state $A$ enters its successor state $B$: $A \to B$, for which $S(t_A) < S(t_B)$ and hence $t_A < t_B$. 

The main problem is to choose an appropriate physical system representing states $A$. Natural proposal is that such a system does not interact with the surroundings and the all kind of the entropy contributions should be taken into account. These demands are to strong to be fulfilled exactly. Hence, the system should have well defined the entropy store and experience negligibly small interaction with the environment.

The cosmological arrow of time considers our Universe as such a system, whose specific state is characterized by the scale factor $a(t)$ — a quantity that is monotonically increasing with time according to the Friedmann equation in the standard FRW metric.
Penrose introduced the concept of gravitational entropy, which is proportional to the Weyl curvature. The latter grows with the volume of the Universe \cite{Penrose1979}. Building on this relation, the entropy of non-gravitating states was elaborated in \cite{Lieb_1999}. Consequently, entropy is a monotonically increasing function of time, and the application of the Postulate of Causality immediately leads to the arrow of time. Nevertheless, some problems remain. It is unclear how the initial state acquired its extremely low entropy, and why the time direction is the same in different spatial regions. There are debates on the gravitational entropy properties locally, at the spherical collapse \cite{Chakraborty:2024rpi}.

A central thesis of this letter is to address the following questions: What kind of physical states, generally referenced by the Causal Postulate, lead to the emergence of the arrow of time? Does a temporal direction exist at each point in space even in the absence of matter? Can time change its direction in different spatial areas?

It is shown here that applying the Causal Postulate to extra dimensions yields promising results, suggesting that the dominant portion of entropy is stored within these additional dimensions.

We do not consider the subtle question of whether the future is static or arising, which is substantially discussed in \cite{Ellis:2006sq}. A third possibility—that reality has a temporal structure describing becoming—is discussed in \cite{Rovelli:2019igm}. In this article, we focus solely on the mechanism that drives the temporal coordinate forward in an unchanging direction.

\subsection{Choice of physical system}

The first necessary step is to choose a physical realization for the abstract state A in the postulate. To illustrate this, we recall the essence of the cosmological arrow of time and examine how the postulate operates in this well-known case. Specifically, we consider this phenomenon during the inflationary stage.
In quasi-de Sitter inflation, the scale factor grows exponentially:
\begin{equation}\label{at}
a(t) = a_0 e^{H t}, \quad H \approx \text{const.},
\end{equation}
where $H$ is the Hubble parameter. This function is monotonically growing function of the temporal coordinate $t$.
The physical volume $V_{\text{phys}}(t)$ containing a fixed comoving volume $V_c$ thus expands as:
\begin{equation}
V_{\text{phys}}(t) = a(t)^3 V_c \propto e^{3H t}.
\label{eq:physical_volume}
\end{equation}

The 3D space under the de Sitter horizon is a causal patch of constant physical radius $R_H = H^{-1}$ (setting $c=1$) and volume $V_{\text{hor}} \sim H^{-3}$. The number of such independent Hubble patches ${N_4}(t)$ within the growing physical volume is:
\begin{equation}
{N_4}(t) = \frac{V_{\text{phys}}(t)}{V_{\text{hor}}} \propto e^{3H t}.
\label{N_4}
\end{equation}
This exponential increase in the number of independent degrees of freedom is the key to understanding entropy growth. The specific form of the statistical weight for a gravitational system is determined by its geometric properties, which allows us to utilize the Bekenstein-Hawking-Wald formalism. 
For a de Sitter metric, the Bekenstein-Hawking entropy \cite{Bekenstein:1973ur,Hawking:1975vc,Gibbons:1977mu,Balasubramanian:2025hns} is almost constant:
\begin{equation}
S_{\text{hor}} = \frac{ A}{4 L_{\text{Pl}}^2} = \frac{ \pi}{H^2 L_{\text{Pl}}^2} \quad (k_B=1),
\label{eq:horizon_entropy}
\end{equation}
where $A = 4\pi H^{-2}$ is the horizon area and $L_{\text{Pl}}$ is the Planck length.

The total entropy $S_{\text{total}}(t)$ in the physical volume can be estimated as the sum of contributions from all causally independent patches. Since the patches are independent, the total state count multiplies, leading to an additive entropy:
\begin{equation}\label{4entropy}
S_{\text{total}}(t) \approx S_{\text{hor}} \times {N_4}(t) \propto S_{\text{hor}} \cdot e^{3H t}.
\end{equation}
The common consensus holds that entropy increases with time. However, this observation does not, in itself, explain the emergence of the arrow of time, as it remains unclear why a system would transition from one state to a subsequent state of higher entropy.  This gap in the classical framework necessitates a Postulate of Causality, which cannot be derived from existing physical laws but must be introduced as a fundamental principle. 

The Postulate of Causality declares a transition $M(t_0 )\to M(t_1)$ for which the entropy satisfies the inequality $S(t_0)<S(t_1)$. As shown in \eqref{tS}, the inflation supplies us with monotonic function $t(S)$, so that 
\begin{equation}\label{Stot}
S(t_0)<S(t_1)\Longrightarrow t_0<t_1    .
\end{equation}
This process is interpreted as the monotonic increase of the temporal coordinate $t$, thereby establishing the arrow of time. Specifically, by inverting the entropy growth relation in \eqref{4entropy}, we obtain:
\begin{equation}\label{tS}
t \propto \ln S_{\text{total}}
\end{equation}
which formally demonstrates that the flow of time is a logarithmic function of the total geometric entropy.

In this subsection, we examine the viability of the Postulate of Causality in establishing an arrow of time during the inflationary stage. The central inquiry of this study, however, concerns the local arrow of time in the current epoch. This leads to a fundamental question: does an arrow of time persist even in a vacuum? Defining the entropy within a volume becomes problematic when matter is absent or limited to a few neutral particles. Nevertheless, the arrow of time appears to remain robust. As shown below, the introduction of extra dimensions provides a consistent framework for relating time and entropy, even in the total absence of 3D matter.

\section{The multidimensional arrow of time}

Extra dimensions play a significant role in modern theoretical physics. If this concept is valid, it is worthwhile to examine its influence on the entropy evolution.

Here, we have shown that the multidimensional arrow of time can be treated in the same manner as the The Cosmological arrow of time. 
The system acting in the Postulate of Causality is changed - the 3D manifold describing our Universe is substituted by the $D-1$-dimensional space. We have to find monotonicaly growing solution like \eqref{at}, endow the extra dimensions by an entropy and show that the entropy is proportional to the temporal coordinate similar to \eqref{tS}.

For the subsequent analysis, we utilize the action:
\begin{equation}   \label{S}
    S_g =   \frac{m_D^{D-2}}{2}\int d^D x \sqrt{|g_D|}  f(R) 
\end{equation}
where $g_D = \det (g_{AB})$,  $f(R)$ is some function of the D-dimensional scalar curvature $R$. Variation of  $S_g$ with respect to $g^{AB}$ leads to the field equations in the general form
\begin{equation}\label{fReq} -\frac{1}{2} \delta_A^B f(R) + \left[ R_A^B + \nabla_A \nabla^B - \delta_A^B \Box \right] f_R = 0, \end{equation} 
where $f_R = df/dR$ 

The metric used in the remainder of this article takes the following form:
\begin{equation}\label{genmet}
ds^2=e^{\gamma(t,u)}(dt^2-e^{2Ht}d\vec{x}^2)-e^{\beta(t,u)}(du^2 + r^2(t,u)d\Omega_{n-1}^2)
\end{equation}
We are interested in two types of asymptotic metrics within the same coordinate system $(t, x_i, u, \Omega_{n-1})$, both of which are solutions to equations \eqref{fReq}. One such case is discussed in this section. Specifically, we perform the analysis in $D=4+n$ dimensions using a standard and widely used asymptotic metric:
\begin{equation}\label{dS}
ds^2 = dt^2 - e^{2Ht} d\vec{x}^2 - e^{2H_et}d\Omega_{n}^2,
\end{equation}
a configuration extensively discussed in the literature (see, e.g., \cite{2000PThPh.103..893O,Nieto:2004cd,Lyakhova:2018zsr,Pavluchenko:2021eym,Anchordoqui:2023etp}), $n=1,2,...,$ is the number of extra dimensions. %The temporal coordinate $t$ retains its standard physical meaning in describing the evolution of the system. 
Since the functional form of classical equations and their corresponding solutions is independent of the existence of a formal arrow of time, we can utilize solution \eqref{dS} to system \eqref{fReq} as a metric distribution over the temporal coordinate $t$ to investigate the emergence of thermodynamic directionality.

In the following, we assume $H_e = H$ with known analytical solution for arbitrary function $f(R)$, \cite{Rubin:2020kiy}
$$H=\sqrt{\frac{f(R)}{2(D-1)f_R(R)}}$$
where the Ricci scalar $R = D(D-1)H^2$. It is assumed that conditions $f_R >0,\, f_{RR}>0$ holds.
 
The spatial volume of the $(D-1)$-dimensional hypersurface depends on the temporal coordinate as:
$$V_{D-1}\propto e^{(D-1)Ht}$$
which is standard for the $D$-dim de Sitter metric.

To characterize the thermodynamics of this $D$-dimensional manifold, we employ the Wald entropy framework. For a gravitational theory described by a general action with an arbitrary function $f(R)$, the entropy is defined via the Noether charge associated with the horizon  \cite{Wald:1993nt, Nojiri:2010wj, Volovik:2024dec}:
\begin{equation}
S_{Wald} = \frac{f_R(R) A_{D-2}}{4 G_D} = \frac{f_R(R) A_{D-2}}{4 L_D^{D-2}}
\end{equation}
where $L_D$ is the $D$-dimensional Planck length, and the area of the $(D-2)$-dimensional sphere of radius $1/H$ is given by:
\begin{equation}
A_{D-2} = \frac{2\pi^{\frac{D-1}{2}}}{\Gamma\left(\frac{D-1}{2}\right)} \left( \frac{1}{H} \right)^{D-2}
\end{equation}
is the horizon area.

It is important to note that the total entropy growth is not limited by the fixed size of a single causal horizon. As the multidimensional space expands, the initial volume increases, leading to the formation of a vast number of causally disconnected regions, each possessing its own horizon. The total entropy of the system is the sum of the entropies of all such regions. Therefore, even if the entropy within a single horizon remains nearly constant, the total entropy $S_{total}$ grows proportionally to the number of these regions
\begin{equation} \label{SWald_Total}
    S^{(D)}_{\text{total}}(t) = \sum_{i} S_{\text{Wald}, i} \approx N_D(t) \cdot \frac{f_R(R) \mathcal{A}_{D-2}}{4L_D^{D-2}}
\end{equation}
By accounting for the exponential growth of the number $N_D$ of causally disconnected regions, we find that the total integrated entropy scales directly with the expanding spatial volume:
\begin{equation}\label{S_D}
    S^{(D)}_{\text{total}}(t) \propto  N_D(t) \simeq\frac{V_{D-1}}{H^{D-1}}\propto e^{(D-1)Ht}   % f_R(R) e^{(D-1)Ht}
\end{equation}
This formulation provides the thermodynamic justification for the arrow of time, as the temporal coordinate  $t$ increases monotonically with the Wald entropy
\begin{equation}\label{tSD}
    t\propto \ln{ S^{(D)}_{\text{total}}}
\end{equation}
providing the physical basis for the Postulate of Causality. The latter is applied in this case in the same manner as in 4D case, see \eqref{Stot}.
We conclude that the multidimensional arrow of time presents as the result of $D-1$ space expansion.

Comparison of the entropy growth rates
\begin{equation}\label{ratio}
    \frac{\dot{S}^{(D)}_{\text{total}}}{\dot{S}_{\text{total}}}\propto e^{nHt}
\end{equation}
indicates that the multidimensional entropy production rate dominates the total entropy budget.

\section{Why don't we see large extra dimension? Branes.}\label{brane}

In the previous sections, we described a possible pathway toward the formation of the arrow of time. The price for this is the existence of continuously increasing extra dimensions. Why then do we not recognize their presence? An answer is provided by the brane-world scenario \cite{Rubakov:1983bb}, which supposes that matter is confined to a brane and does not spread into the bulk. The confinement of the Standard Model fields to a 3-dimensional spatial hypersurface (the 3-brane) ensures that the expansion of the extra-dimensional volume $V_n(t)$—the bulk—does not lead to a direct change in atomic scales.

An observer residing on such a brane would perceive only a small part of the extra space. Therefore, a wider class of solutions to the nonlinear system \eqref{fReq} for the metric functions \eqref{genmet} should describe a brane in the local vicinity of some extra coordinate $u=u_b$, which asymptotically tends toward the maximally symmetric extra space with the growing radius described above. Such solutions are too complex to be found in their complete form. Nevertheless, it is sufficient for our purpose to prove that the static brane solutions for the same function $f(R)$ do exist.

We choose brane-like metric a lot examples of which had been studied, see e.g.  
\cite{Maartens:1999hf,Peyravi:2015bra}, and review \cite{Brax:2003review}. It is also worth mentioning the thin shell approach considering the dynamic of the spherical bubbles \cite{ColemanDeLuccia1980,Berezin:1987bc} surrounded by another vacuum. The stability of the two-brane metric is discussed in \cite{Burgess:2001bn}. The branes are separated from $D-1$ space expansion by definition. An observer located in the brane feels physics depending on the brane metric. As the example, we consider the two-brane solution with the warped inhomogeneous metric of the following form, \cite{Bronnikov:2023lej}:
\begin{equation}\label{metricb}
ds^2 = e^{2\gamma(u)} \left( dt^2 - e^{2H_bt} d\vec{x}^2 \right) -  \left( du^2 + r(u)^2 d\Omega_{n-1}^2 \right),
\end{equation}
where $\gamma(u)$ is the conformal warp factor, $H_b$ is the constant Hubble parameter of the 3D space, and $r(u)$ is the radius of the extra subspace.
Note that the temporal coordinates $t$ in both metric \eqref{dS} and \eqref{metricb} are equal. In the previous section we show that the arrow of time $t$ exists due to the space expansion of the bulk.
As was shown in \cite{Maartens:1999hf,Bronnikov:2023lej} the brane metric \eqref{metricb} is able to restore the inflationary scenario including the reheating stage, the Starobinski model in particular and leads to the observable effects \cite{Rubin2025}.  In \cite{Popov:2024nax} it was shown that scalar, fermionic and gauge fields are located on branes. 

The scale of branes is limited by the $D$-dim horizon. Such branes - solutions to classical equations \eqref{fReq} with the same action \eqref{S} describing the branes \eqref{metricb} and the expanding extra dimensions \eqref{dS} have been studied separately, but with the same function of $f(R)$ gravity and the dimensionality of extra space, in \cite{Rubin:2020kiy} and \cite{Bronnikov:2023lej}. The similar research can be found in \cite{Zhong_2016}, \cite{Guo:2024izl}. This approach is sufficient for our purpose, albeit at the cost of losing a smooth transition between the two regimes. 
The analogy follows the standard practice in cosmology: if our aim is to study the structure of a galaxy, we neglect processes like spatial expansion near the galaxy and solve the equations using the Minkowski metric. To study cosmological expansion, we consider the FRW metric as the solution to the Friedmann equations far from any local object. Matching these solutions at the boundary is of special interest and is not usually applied. 

Let us address a reasonable question that arises at this stage. At first sight, the arrow of time appears to emerge from the global expansion far from the horizon. Conversely, an observer within the horizon cannot, by definition, perceive the total expansion. Why, then, is the arrow of time still recognized by such an observer? 

To clarify this situation, we consider a similar problem within the framework of the McVittie metric, which describes a black hole of mass $m$ embedded in a $D$-dimensional de Sitter background. Its primary advantage for our purposes is that its line element is well-known and possesses a closed analytical form

\begin{equation}
ds^2 = \left( \frac{1 - \frac{m}{4 a(t)^{D-3} r^{D-3}}}{1 + \frac{m}{4 a(t)^{D-3} r^{D-3}}} \right)^2 dt^2 - a(t)^2 \left( 1 + \frac{m}{4 a(t)^{D-3} r^{D-3}} \right)^{\frac{4}{D-3}} (dr^2 + r^2 d\Omega^2_{D-2}),
\end{equation}
where $m$ represents the mass of the static compact object (e.g., a star or a black hole) and $a(t) = e^{Ht}$ is the scale factor of the background universe. To determine the region of gravitational dominance, we look for the radius $R_{c}$ where the inward gravitational pull exactly balances the outward de Sitter expansion. In $D$ dimensions, the physical radius of this "static shell" is given by:
\begin{equation}
R_{c} = \left( \frac{(D-3)m}{2H^2} \right)^{\frac{1}{D-1}}
\end{equation}
For any observer located at a physical distance $R(t) < R_{static}$, the geometric state of the manifold allows for traditional gravitational attraction toward the central mass $m$.

Therefore, in this single coordinate system, two distinct spatial regimes coexist. While the positive cosmological constant drives accelerated expansion on cosmological scales, its influence remains negligible in the strong-field region near the compact object. Consider an observer at rest at a physical radius $R$. If the observer is positioned below the critical radius, at   $R < R_{crit}$, they are effectively decoupled from the global expansion. In this regime, the proper time $d\tau$ relates to the cosmological time $dt$ as:
\begin{equation}
    d\tau = \frac{1 - \frac{m}{2r^{D-3}a(t)^{D-3}}}{1 + \frac{m}{2r^{D-3}a(t)^{D-3}}} dt
\end{equation}
The factor in the r.h.s. is a monotonic function of both $t$ and $r$. Consequently, if $dt > 0$, then $d\tau > 0$; thus, the local observer experiences a well-defined proper arrow of time.

In our model, the brane plays a role analogous to this decoupled region: the brane remains a stable, static entity residing deep ``under the horizon,'' while the global bulk continues its expansion. This configuration ensures that the established arrow of time is maintained on the brane despite its local stability.

The Past Hypothesis postulates that the initial entropy of the Universe was extremely small compared to the entropy within the cosmological horizon today, \cite{Kiefer:2009xr}. Let us estimate their ratio within the framework of this work, using equation \eqref{ratio}:
\begin{equation}
    \xi = \frac{S^{(D)}(t)}{S^{(D)}(0)} \sim \frac{N_D(t)}{N_D(0)} \sim e^{(D-1) H t}
\end{equation}
Here,  \( S^{(D)}(0) \)  denotes the Wald entropy contained within the cosmological horizon at the moment $t=0$ when our brane was created.  \( S^{(D)}(t) \) represents the Wald entropy of the same comoving volume at the present cosmic time, $t$.

For a quantitative estimate, we set $D = 6$, use the Hubble parameter during inflation $H \sim 10^{14} \, \text{GeV} \sim 10^{38} \, \text{s}^{-1}$, and take the lifetime of the Universe $t \sim 10^{17} \, \text{s}$. This yields an enormous ratio:
\begin{equation}\label{xi}
  \xi \sim 10^{10^{55}}.  
\end{equation}

The entropy $S_D$, defined in Eq. \eqref{S_D}, exhibits permanent growth in the multidimensional space. This constant flow of entropy dominates over any local entropy changes on the brane and supplies the arrow of time permanently.  The initial value is extremely small compared to the present time and requires no fine-tuning within the framework of our model.

To highlight the scale of this result, it is instructive to compare the calculated geometric entropy with the entropy of matter in the observable Universe. The total entropy of matter and radiation is estimated to be approximately $10^{90} - 10^{100}$, while the entropy of the cosmic event horizon in the standard 4D $\Lambda$CDM model is about $10^{122}$. In contrast, the entropy growth factor \eqref{xi} obtained in our model, is incomparably larger. This colossal discrepancy demonstrates that the thermodynamic arrow of time is not merely a consequence of matter distribution or local fluctuations. Instead, it is a fundamental geometric phenomenon driven by the expansion of the multidimensional bulk. The geometric entropy acts as an inexhaustible 'source,' ensuring that the probability of a global time reversal is effectively zero, making the arrow of time a robust, universal property of the manifold itself.

The general picture is as follows: the arrow of time persists, driven by the entropy growth associated with the multidimensional bulk expansion. Both the brane and the surrounding bulk metrics are described using a common temporal coordinate $t$. While the expansion of the multidimensional bulk remains unobservable due to matter confinement, it determines the global entropic gradient of the $D$-dimensional manifold. For a brane-localized observer, the arrow of time for any local process—such as the motion of a particle—is thus a consequence of the system's evolution toward a state of higher total entropy, dominated by the growth of the multidimensional volume.

A crucial feature of a multidimensional arrow is that it is robust.
Even if our specific pocket universe were to undergo a "Big Crunch" (where the local cosmological arrow might theoretically stall or appear to reverse), the multidimensional arrow remains unaffected. The global process of "geometric nucleation" at scales much smaller than $10^{-27}$ cm ensures that the multiverse as a whole never reaches equilibrium.

\section{Conclusion}

This paper investigates the role of extra dimensions in the emergence of the arrow of time. As is well known, an initial volume, defined in a space with a $D$-dimensional de Sitter metric, grows continuously, generating disconnected volumes bounded by horizons. Each such volume is endowed with the Bekenstein-Hawking-Wald entropy. Consequently, the total entropy of the initial volume increases exponentially along the temporal coordinate.

Our calculations for a 6D model yield an entropy growth factor of $\xi \sim 10^{10^{55}}$ over the lifetime of the Universe. Comparing this result with the maximum entropy of matter in our Universe ($\sim 10^{100}$), it becomes evident that the geometric entropy reservoir is the dominant factor in the overall balance.

Observers are located on a static brane with a scale significantly smaller than the de Sitter horizon. Since matter is, by definition, confined to the brane, the bulk expansion remains unobservable to a local observer. However, the entropy growth of the extra dimensions provides a rigorous basis for the arrow of time. To this end, following Penrose's idea, we introduce a Postulate of Causality related to the entropy of physical states. We demonstrate that the arrow of time arises from the fact that the multidimensional volume is a monotonic function of the temporal coordinate. Conversely, the observer's proper time  $\tau$ is a monotonic function of entropy \eqref{tSD}; therefore, applying the Postulate of Causality establishes the arrow of time.

The advantage of this approach is that it obviates the need for an extremely low-entropy initial state, as the entropy stored in the extra dimensions grows continuously. Furthermore, from the perspective of a 4D observer, the arrow of time persists even in empty space. The immense reservoir of entropy in the extra dimensions dominates over any local entropy fluctuations, making a global reversal of the arrow of time statistically impossible. Notably, such reverse processes are also precluded by the structure of both classical and Schrödinger equations. Although these equations are formally time-reversal symmetric, they exclude "temporal zigzags" \cite{Rubin:2022biz}, thereby ensuring the global consistency of the chosen causal direction of time.

Finally, let us distinguish between the cosmological arrow of time and the multidimensional one. The cosmological arrow of time originated during the inflationary stage within a horizon of $1/H \approx 10^{-27}$ cm and has since physically expanded to $1/H \approx 10^{28}$ cm at the present epoch. Its total entropy is stored in matter and, most significantly, within black holes. 

In contrast, the multidimensional arrow of time is nucleated at much smaller scales, making its physical extent many orders of magnitude larger than the size of our observable universe. Its entropy, governed by the Bekenstein-Hawking-Wald framework, is stored within a constantly increasing number of pocket universes, each residing beneath its own causal horizon.

\section*{Acknowledgment}
The author is grateful to V. Berezin and A. Galiautdinov for their interest in the research.
The work was partly funded by the Ministry of Science and Higher Education of the Russian Federation, Project "Studying physical phenomena in the micro- and macro-world to develop future technologies"\, FSWU-2026-0010
and the Kazan Federal University Strategic Academic Leadership Program. 

\bibliographystyle{unsrturl}
\bibliography{Ru-Article_7.bib}

\end{document}